\documentclass[12pt,preprint]{aastex}
\begin{document}

\title{Re-analysis of VLT Data for M83 with Image Subtraction ---
Nine-fold Increase in Number of Cepheids}

\author{A. Z. Bonanos, K. Z. Stanek}
\affil{Harvard-Smithsonian Center for Astrophysics, 60 Garden St.,
Cambridge, MA~02138}
\affil{\tt e-mail: abonanos@cfa.harvard.edu, kstanek@cfa.harvard.edu}

\begin{abstract}

We apply the image subtraction method to re-analyze the ESO Very Large
Telescope data on M83 (NGC 5236), obtained and analyzed by Thim et
al. Whereas Thim et al. found 12 Cepheids with periods between 12-55
days, we find 112 Cepheids with periods ranging from 7-91 days, as
well as $\sim60$ other variables. These include 2 candidate eclipsing
binaries, which, if confirmed, would be the first optically discovered
outside the Local Group. We thus demonstrate that the image
subtraction method is much more powerful for detecting variability,
especially in crowded fields. However, {\em HST}\/ observations are
necessary to obtain a Cepheid period-luminosity distance not dominated
by blending and crowding.  We propose a ``hybrid'' approach, where
numerous Cepheids are discovered and characterized using large
ground-based telescopes and then followed-up with the {\em HST}\/ to
obtain precise distances.

\end{abstract}
\section{Introduction}

Cepheid variables are one of the most important primary distance
indicators due to the period-luminosity (PL) relation they obey.
Precise Cepheid distances require a large number of Cepheids with
accurate photometry and periods. However, the more distant the galaxy,
the more important become the effects of crowding and blending with
nearby luminous stars (Mochejska et al. 2000; Stanek \& Udalski 1999;
Mochejska et al. 2001), which introduce systematic, one-sided errors
in the distance measurements and skew the distance towards
artificially lower values. These effects can only be disentangled
satisfactorily with the {\em Hubble Space Telescope}\/, which has the
required spatial resolution, especially now with the new
instrumentation (ACS).

The spiral galaxy M83, the principal member of the nearby M83 group,
is a good target to obtain a Cepheid distance to. Observations of M83
with the Very Large Telescope (VLT) were analyzed by \citet{Thi03}
with point-spread function (PSF) photometry, discovering 12 Cepheids,
which were used to derive the Cepheid distance to the galaxy. However,
the ``traditional'' method of doing PSF fitting photometry on all the
stars on an image and looking for variations in the light from night
to night is not very efficient or effective in finding Cepheids in
crowded fields. In this paper we re-analyze the VLT data for M83 using
image subtraction to demonstrate this fact.

Observations of M83, at a distance of $\sim4.5$ Mpc \citep{Thi03},
with the 8.2 meter VLT and $0.76\arcsec$ median seeing are roughly
equivalent to the DIRECT project observations of M31/M33 with the FLWO
1.2 meter telescope, at a distance of $\sim780$ kpc (Stanek \&
Garnavich 1998) with $\sim4\arcsec$ seeing. This fact partly motivated
this paper -- we would consider such poor seeing data unacceptable to
run PSF photometry on. The three most recent DIRECT papers
\citep{Moc01a,Moc01b, Bon03} have used Alard's image subtraction
package ISIS \citep{Ala98,Ala00} to discover variables, in particular
detached eclipsing binaries and Cepheids, in M31 and M33. This method
has become a method of choice for variability searches in crowded
fields.

At the distance of M83, the issue of blending must be taken into
account in deriving the Cepheid distance. The median seeing of the VLT
data is $0.76\arcsec$, which corresponds to $17\;$pc in M83. As first
discussed by Mochejska et al. (2000), blending is the close
association of a Cepheid with one or more intrinsically luminous
stars, which is the result of the higher value of the star-star
correlation function for massive stars, such as Cepheids, compared to
random field stars.  This effect cannot be detected within the
observed PSF by usual analysis. \citet{Moc00,Moc01c} address the
effect of blending on the Cepheid distances to M31 and M33. They
compare high resolution {\em HST}\/ images to the ground-based DIRECT
data and find that blending can affect the flux of a Cepheid typically
by $\sim20-30\%$, which leads to an underestimation of the true
distance. This phenomenon is different from crowding, which is the
random background luminosity fluctuation in each resolution
element. In M83, a large fraction of the flux of a blended Cepheid
could come from its companions, assuming the effect scales to that in
M31 and M33. Based on a simulation by Stanek \& Udalski (1999), this
would result in a significant distance bias.  Whereas the discovery of
Cepheids in nearby galaxies can be done adequately from the ground
given good signal-to-noise photometry, deriving the Cepheid PL
distance requires high spatial resolution {\em HST}\/ imaging.

In this paper, we apply the image subtraction method to the M83
dataset. In Section 2 we describe the observations, in Section 3 the
image subtraction method and finally the results, which are discussed
in Section 4.

\section{Observations}

We have retrieved the M83 data obtained by FORS1 on the ESO Very Large
Telescope, from the ESO/ST-ECF Science Archive Facility. The field of
view is $6.8\times6.8$ arcminutes, with a pixel scale of $0.2\arcsec$
per pixel. The $2048\times2048$ Tektronix CCD has 24 $\mu$m
pixels. There are 34 epochs in $V$-band spanning a period of 1.5
years, from January 2000 to July 2001. Each epoch consists of 3-4
subexposures of 400 or 500 sec. More details are given by
\citet{Thi03}.
\vspace{-0.5cm}
\section{The Method of Image Subtraction and Results}
The images were overscan corrected and flat fielded with standard
IRAF\footnote{IRAF is distributed by the National Optical Astronomy
Observatories, which are operated by the Association of Universities
for Research in Astronomy, Inc., under cooperative agreement with the
NSF.} routines. The transformation from rectangular to equatorial
coordinates was derived using 84 transformation stars from the
USNO-B1.0 \citep{Mon03} catalog. The average difference between the
catalog and the computed coordinates for the transformation stars was
less than $0.\arcsec3$ in RA and $0.\arcsec3$ in Dec. Next, we ran the
image subtraction package ISIS \citep{Ala98,Ala00} on the 104 $V$-band
images of M83.

The ISIS reduction procedure consists of several steps. Initially, all
the frames are transformed to a common coordinate grid. The best
seeing frame, FORS.2000-04-05T06:16:02.924 in our case, is chosen to
be the reference image. Next, a composite reference image is created
by stacking several best seeing frames. For each frame, the composite
reference image is convolved with a kernel to match its PSF and then
subtracted. On the subtracted images, the constant stars cancel out,
and only the signal from variable stars remains. A median image is
constructed of all the subtracted images, and the variable stars are
identified visually as significant peaks. Finally, profile photometry
is extracted from the subtracted images. \citet{Moc01a} describes this
procedure in more detail.

\begin{figure}   
\plotone{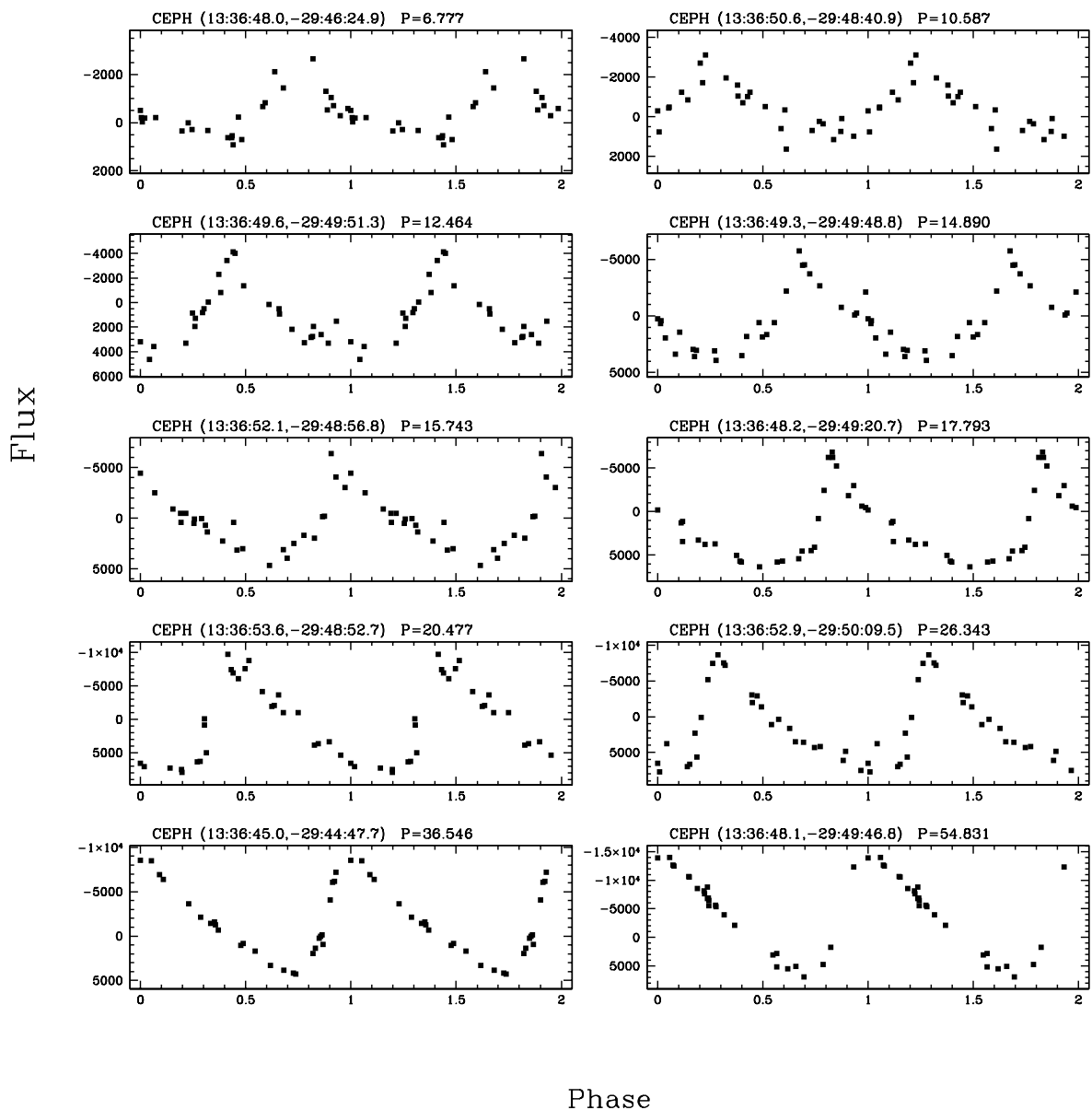}
\caption{Flux light curves of 10 sample Cepheids found in M83. For
each Cepheid, the coordinates (RA, Dec) are given in parentheses,
along with the period. The median point for each epoch is plotted.}
\label{CephLC}
\end{figure}

\begin{figure}   
\plotone{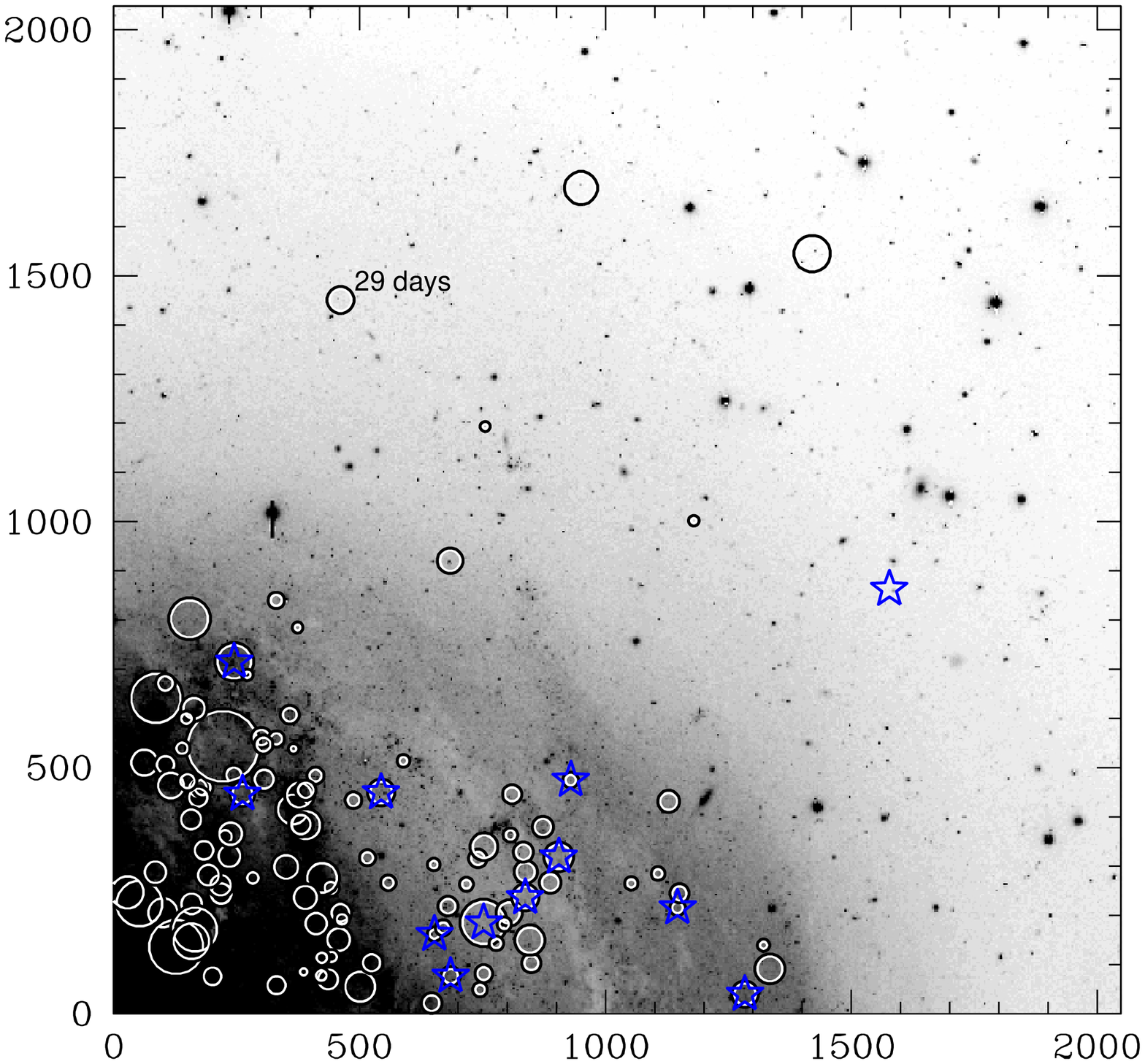}
\caption{Location of the 112 Cepheids we found with image subtraction
on the reference image. Circles are centered at the position of the
Cepheids and their size is proportional to the period of the Cepheid
(see labeled 29 day Cepheid for reference). Stars indicate the
positions of the 12 Cepheids found by \citet{Thi03}. All but one were
recovered by our analysis. [{\it See the electronic edition of the
Journal for a color version of this figure.}]}
\label{CephDist}
\end{figure}

After locating the positions of the bright peaks, we obtained light
curves for $\sim650$ candidate variables. We ran them through the
DIRECT pipeline \citep{Kal98,Sta98} which fits model Cepheid light
curves and eclipsing binary light curves and classifies the stars as
Cepheids, eclipsing binaries (EBs) or other variables. After checking
these light curves, we found 112 Cepheids, with periods ranging from 7
to 91 days, 2 candidate EBs and $\sim60$ other variable
stars. Figure~\ref{CephLC} presents 10 sample Cepheid flux light
curves. Table~\ref{tab:ceph} lists the coordinates (RA, Dec) and (X,
Y) on the reference image for each Cepheid and the periods. We recover
all but one of the Cepheids found by \citet{Thi03}, the Cepheid
C4. The periods we derive agree well with those of \citet{Thi03}, the
median difference being 0.04 days and the largest difference being
0.22 days for Cepheid C6. Figure~\ref{CephDist} plots the location of
the Cepheids as circles on the reference image, their size being
proportional to the period. A 29 day Cepheid is labeled for scale. On
the same plot, the stars show the position of the 12 Cepheids from
\citet{Thi03}. The image subtraction method detects many Cepheids in
the more crowded part of the field, which is much harder to do with
the ``traditional'' PSF photometry method.

We do not claim completeness in finding Cepheids, however we do show
that Cepheids in the highly crowded and blended central regions of the
galaxy can be detected from the ground with high quality data, such as
these VLT observations. High resolution observations are necessary to
disentangle blending for Cepheids both in the outer and inner parts of
the galaxy.
\vspace{-0.5cm}
\section{Discussion}

This paper re-analyzes the excellent VLT data of M83 obtained by
\citet{Thi03} using the image subtraction method. The resulting
nine-fold increase in the number of Cepheids detected indicates that
image subtraction should be used in crowded fields. We also present
parameters and sample light curves for the 112 Cepheids we have
found. These additional Cepheids are valuable for determining the PL
distance to M83 accurately. However, {\em HST}\/ observations are
necessary to resolve blending effects. After we started working on
this project, we became aware of the Cycle 12 program with the title
``M83: Calibrating the Cepheid PL Relation'' (PI: B. Madore). We note
that some of these 112 Cepheids might be Population II variables,
however, a PL diagram would be necessary to distinguish them.

\begin{figure}   
\plotone{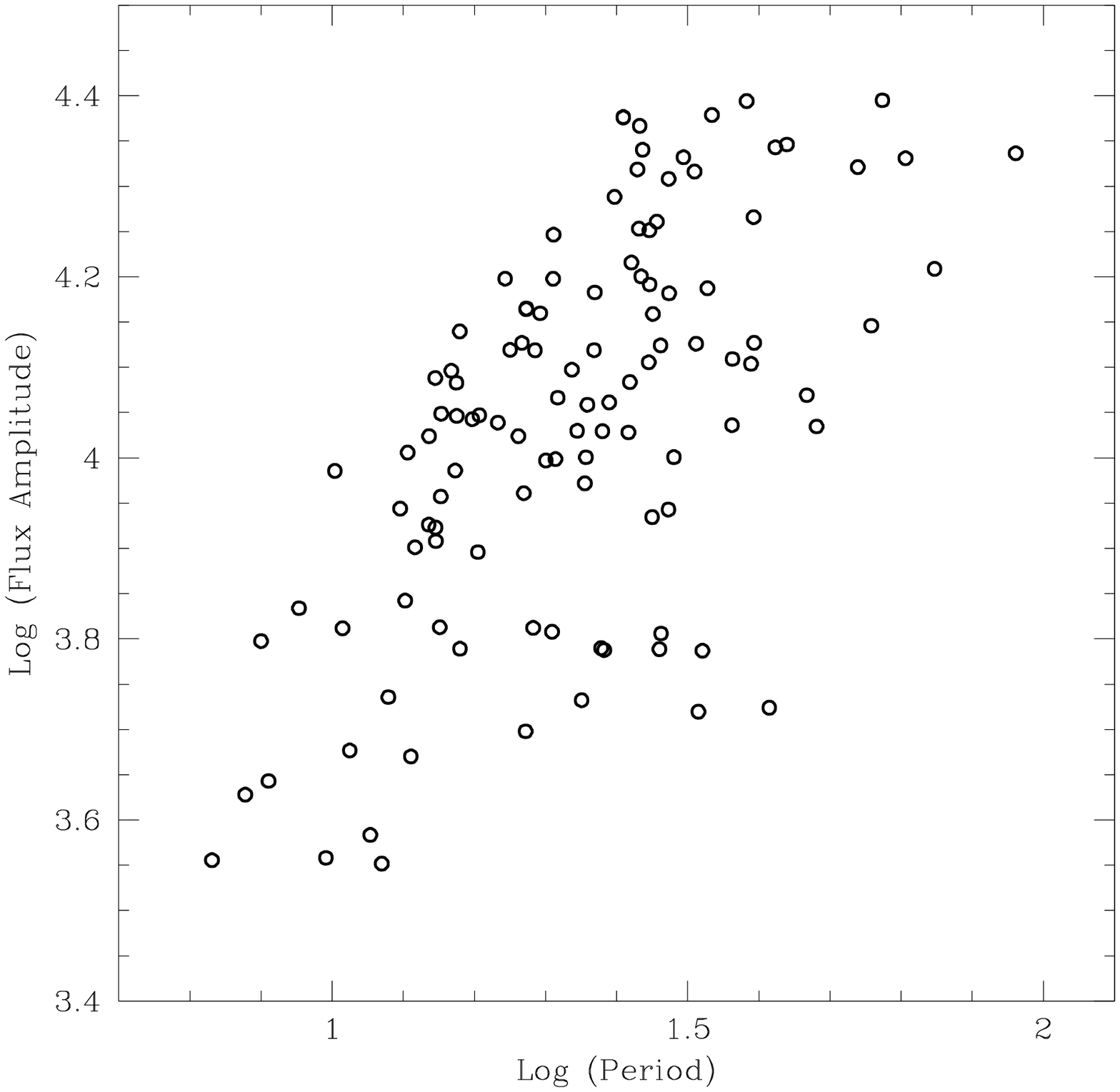}
\caption{Flux amplitude versus period for 112 Cepheids in M83.}
\label{Ampl}
\end{figure}

In Figure~\ref{Ampl}, we plot the flux amplitude versus period
relation for the 112 Cepheids. We have taken the full flux amplitude
from the light curve and plotted the logarithm of this quantity versus
logarithm of the period $P$. There is a definite correlation, however,
it is not very tight. \citet{Pac00} demonstrate that the period-flux
amplitude relation is not universal and needs to be calibrated before
being used to measure distances accurately.

\begin{figure}   
\plotone{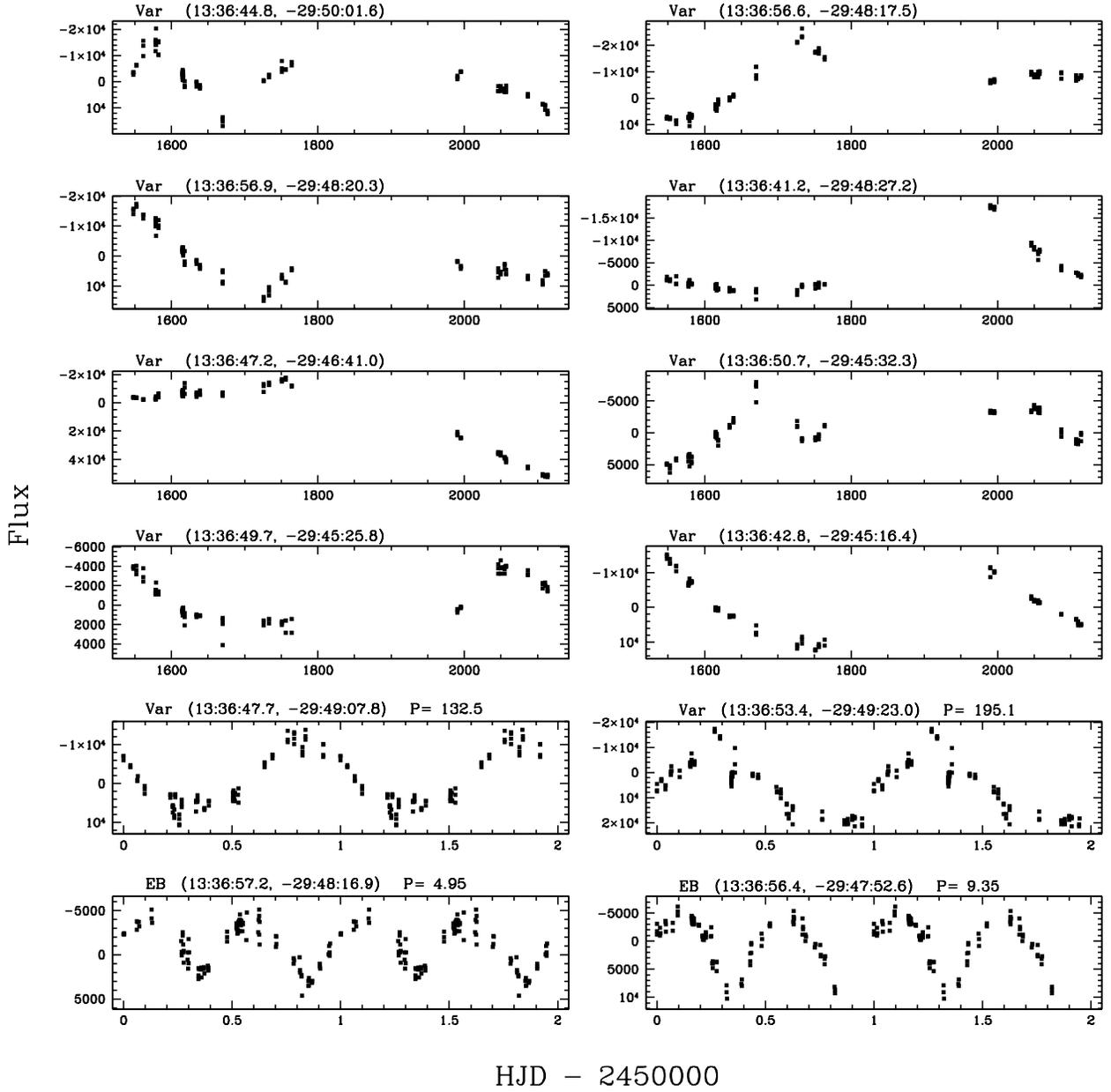}
\caption{Flux light curves of some miscellaneous variables in M83,
with coordinates (RA, Dec). Two possible long period Cepheids and two
candidate eclipsing binaries are shown.}
\label{Misc}
\end{figure}

In addition to these Cepheids, we also find $\sim60$ other variable
stars, including 2 candidate eclipsing binaries (EBs). If these are
confirmed from their position on a CMD to be located in M83, they
would be the first optically discovered EBs outside the Local
Group. Figure~\ref{Misc} presents some of the more interesting light
curves of other periodic or non periodic variables that we find,
including the candidate EBs. There are two long period variables,
which are possibly Cepheids with periods of 132 and 195 days. Their
positions on the PL diagram would verify this.

\vspace{-0.5cm}
\acknowledgments{We thank Danny Steeghs, David Bersier and Barbara
Mochejska for helpful discussions. We thank Grzegorz Pojma\'nski for
his most useful ``lc'' program. We also thank Bohdan Paczy\'nski and
Dimitar Sasselov for their comments on an earlier version of this
paper. This work is based on observations made with the European
Southern Observatory telescopes obtained from the ESO/ST-ECF Science
Archive Facility.}

\begin{deluxetable}{lrrrrrl}
\tabletypesize{\footnotesize}
\tablewidth{0pc}
\tablecaption{\sc Cepheids in M83}
\tablehead{
\colhead{} & \colhead{} & \colhead{$P$} & \colhead{} & \colhead{}
&\colhead{}
\\ \colhead{RA} & \colhead{Dec}& \colhead{$(days)$} & \colhead{X} &
\colhead{Y} &\colhead{Comments$^{\rm a}$}} 
\startdata
13:36:50.6 & $-$29:48:40.9  &  10.587 &  589.28 &  514.04 &\\
13:36:43.4 & $-$29:49:31.0  &  11.313 & 1052.49 &  264.76 &\\
13:36:42.6 & $-$29:49:27.0  &  11.745 & 1106.87 &  285.00 &\\
13:36:48.6 & $-$29:49:31.1  &  11.997 &  717.45 &  263.21 &\\
13:36:49.6 & $-$29:49:51.3  &  12.464 &  652.30 &  161.81 & C9, 12.47 days\\
\enddata                         

\label{tab:ceph}
\tablenotetext{a}{Names and periods of Cepheids found by
\citet{Thi03}.}  \tablecomments{Table 1 is available in its entirety
in the electronic version of the Journal. A portion is shown here for
guidance regarding its form and content.}
\end{deluxetable} 

\end{document}